\documentclass[journal=jacsat,manuscript=article]{achemso}

\usepackage{chemformula} 
\usepackage[T1]{fontenc} 

\author{Freiman Triana-Arango}
\email{freiman@cio.mx}
\affiliation[Centro de Investigaciones en Óptica A.C., A. P. 1-948, 37000 León, Gto, Mexico]
{Centro de Investigaciones en Óptica A.C., A. P. 1-948, 37000 León, Gto, Mexico}
\author{Roberto Ramírez-Alarcón}
\email{roberto.ramirez@cio.mx}
\affiliation[Centro de Investigaciones en Óptica A.C., A. P. 1-948, 37000 León, Gto, Mexico]
{Centro de Investigaciones en Óptica A.C., A. P. 1-948, 37000 León, Gto, Mexico}
\author{Gabriel Ramos-Ortiz}
\email{garamoso@cio.mx}
\affiliation[Centro de Investigaciones en Óptica A.C., A. P. 1-948, 37000 León, Gto, Mexico]
{Centro de Investigaciones en Óptica A.C., A. P. 1-948, 37000 León, Gto, Mexico}

\title[An \textsf{achemso} demo]
  {Entangled two-photon absorption in transmission-based experiments: deleterious effects from linear optical losses}

\abbreviations{IR,NMR,UV}
\usepackage{color}
\usepackage{soul}
\usepackage{lineno}
\usepackage{multirow}

\begin{document}

\begin{abstract}
Recently different experimental schemes have been proposed to study the elusive phenomenon of entangled two-photon absorption (ETPA) in nonlinear materials.  The attempts to detect ETPA using transmission-based schemes have led to results whose validity is currently under debate since the ETPA signal can be corrupted or emulated by artifacts associated with linear optical losses. The present work addresses the issue of linear losses and the corresponding artifacts in transmission-based ETPA experiments through a new approach that exploits the properties of a Hong-Ou-Mandel (HOM) interferogram. Here we analyze solutions of Rhodamine B (RhB), commonly used as a model of nonlinear medium in ETPA studies. Then, by using the HOM interferometer as a sensing device, we firstly demonstrate the equivalence of the standard transmission vs pump power ETPA experiments, presented in many reports, with our novel approach of transmission vs two-photon temporal delay. Secondly, a detailed study of the effects of optical losses, unrelated to ETPA, over the HOM interferogram is carried out by: 1) Characterizing RhB in solutions prepared with different solvents. 2) Considering scattering losses introduced by silica nanoparticles used as a controlled linear loss mechanism. Our results clearly expose the deleterious effects of linear optical losses over the ETPA signal when standard transmission experiments are employed and show how, by using the HOM interferogram as a sensing device, it is possible to detect the presence of such losses. Finally, once we showed that the HOM interferogram discriminates properly linear losses, our study also reveals that under the specific experimental conditions considered here, which are the same than those employed in many reported works, the ETPA was not unequivocally detected. 
\end{abstract}

\section{\label{sec:Introduction}Introduction}
Since the discovery and development of the laser\cite{MAIMAN1960}, and subsequently the nonlinear optical effects\cite{Franken1961,Maker1962,Armstrong1962,Bloemberge1982}, an advent of photonic technologies have been aroused  \cite{Shen1976,Chen1986}. Among these, the two-photon absorption (TPA) process, predicted theoretically by M. Goeppert-Mayer in 1931, has become an useful tool in different areas such as bioimaging\cite{WDenk} and micro-fabrication applications\cite{Wei:07}. TPA is a third-order nonlinear optical process\cite{Maker1965,MariacristinaRumi2010} which requires high-intensity excitation sources and nonlinear optical materials with large TPA cross-sections ($\delta_{c}$), in order to be observed. Typically, pulsed lasers are employed as excitation sources to promote TPA since they provide a high density of random photons. However, the prediction of TPA through the use of extremely dim (low intensity) non-classical light \cite{GeaBanacloche1989,Georgiades1995,JanPerina1998}, namely the entangled two-photon absorption (ETPA) effect, triggered a great interest in understanding the interaction dynamics of this non-classical excitation source with non-linear active materials. The interest in detecting ETPA has led to theoretical and experimental proposals whose results nowadays are still debatable and generate intense scientific research. 

Different physical effects and potential applications, without classical equivalent, have been predicted in this new type of interaction, such as: non-monotonic behavior of ETPA signals with photon temporal delay\cite{Fei1997}, virtual state spectroscopy\cite{Saleh1998,KOJIMA2004323,RobertoLeon2013}, induction of forbidden transitions\cite{Muthukrishnan2004}, entangled two-photon spectroscopy\cite{SVOZILIK201854,Schlawin2018,RobertoLeon2019,Mertenskotter:21}, among others. Nevertheless, the effect of most notoriety is, possibly, the linear behavior of the nonlinear absorption rate ($R_{TPA}$) with the photon flux ($\phi$)\cite{Javanainen1990}, contrasting with the classical case in which the trend is quadratic\cite{MariacristinaRumi2010}: $R_{TPA}=\sigma_{e}\phi+\delta_{c}\phi^{2}$, where $\sigma_{e}$ is the ETPA cross-section, that represents, as in the classical case ($\delta_{c}$), the probability of an electronic transition promoted by the TPA process. The linear behavior of $R_{TPA}$ with $\phi$ in molecular samples has led to estimate that $\sigma_{e}\sim10^{-17}[cm^{2}/molecule]$\cite{DongIkLee2007}, whereas in the classical case is well known that $\delta_{c}\sim10^{-50}[cm^{4}s/molecule]$\cite{Sperber1986,Xu:96,Makarov:08}. Thus, one of the most attractive features of ETPA is that the same TPA effects, produced with a high flux of uncorrelated photons (classical excitation), could be achieved by illuminating the nonlinear material with a low flux of correlated photons (non-classical excitation). In practice, correlated photons are produced by quantum processes such as spontaneous parametric down-conversion (SPDC)\cite{GhoshR1986,KwiatPaulG1995}. 

Some initial experimental configurations were presented to study and quantify the ETPA activity,  and other nonlinear interactions involving entangled photons, in a variety of materials by means of transmission\cite{Dayan2005,LeeDongIk2006,HarphamMichaelR2009} and fluorescence\cite{Dayan2004} schemes. At the same time, other relevant theoretical works established some conditions to optimize \cite{AviPeer2005,Dayan2007,YouHao2009} and perform protocols for ETPA experiments\cite{DanielKeefer2022}. Since then, new schemes and techniques based on both transmission\cite{VillabonaMonsalve2017,VillabonaMonsalve2020} and fluorescence\cite{UptonL2013,VillabonaMonsalve2018,VarnavskiOleg2020,EshunAudrey2022} have been proposed to estimate the value of $\sigma_{e}$ in diverse materials. However, some authors have suggested that the  $\sigma_{e}$ values obtained from transmission-based experimental schemes could be overestimated since the effects of linear losses emulating\cite{Mikhaylov2022,BrycePHickam2022} or contaminating\cite{Parzuchowski2021,LandesTiemo2021,LandesTiemo2021_2,CoronaAquinoSamuel2022} the EPTA signal were not properly discriminated. Therefore, it is worth studying in detail how the measurements of the ETPA signal and the $\sigma_{e}$ values estimated in transmission experiments are affected by optical losses.

In a previous work from our group \cite{Freiman}, we proposed a novel configuration to study the ETPA process by using a HOM interferometer as an ultra-sensitive detection device. In the cited work we presented a theoretical model and implemented the corresponding measurements of transmission vs two-photon temporal delay which allowed us to understand the effects of a nonlinear sample (RhB), modeled as a notch type two-photon spectral filter, over the asymmetric\cite{Grice1997} quantum state of photon pairs produced by type-II SPDC, in the CW and pulsed excitation regimes. In that work, we showed that the ETPA activity can be evidenced as a noticeable change in the HOM dip visibility, produced by the interference of the down-converted photons after the interaction with the sample. Also, we found that the best strategy to observe the ETPA activity in an organic molecule such as RhB, with the HOM interferogram as a sensing device, is to exploit a type-II SPDC source in the pulsed pumping regimen and match the central emission wavelength and bandwidth of the photon pairs to those of the sample under study. Nevertheless, that study did not show the equivalence between the proposed HOM-based transmission vs temporal delay scheme and the standard transmission vs pump power scheme used commonly to measure ETPA, nor did it present a detailed study of the effects produced by lineal optical losses over the detected signal in the HOM dip.   

In the present work, we implement ETPA transmission experiments using RhB as a nonlinear sample under CW excitation and exploiting the capabilities offered by the HOM interferometer as a sensing device, some of them already explored in linear spectroscopy studies\cite{AudreyEshun2021,KonstantinEDorfman2021,YuanyuanChen2022,NicolasFabre2022}.  We choose the CW pumping regime for the SPDC type-II photon pairs source since it is a widely used experimental configuration in reported studies of ETPA activity, for which the detrimental effects of linear losses has never been considered. By analyzing the RhB transmission of photon pairs produced around 800nm, we demonstrate that the scheme of transmission vs two-photon temporal delay, based on the HOM interferogram, is completely equivalent to the standard transmission vs pump power configuration reported by various authors. In addition, a detailed study of the effects of linear optical losses, unrelated to ETPA, is carried out in both schemes. It is shown that while the linear losses lead to overestimating $\sigma_{e}$ when the transmission is detected as a function of pump power, in the case of transmission vs temporal delay the losses only produce a reduction of the raw base level of coincidence counts of a HOM interferogram, but without changing the visibility of its dip. Notice that the visibility of a HOM dip changes upon modifications to the quantum state of the photon pairs, and changes in the quantum state are expected in the case of ETPA activity \cite{Freiman}. Thus, the transmission vs time delay configuration not only is equivalent to the standard scheme used to study ETPA but also offers the remarkable advantage of discriminating the presence of deleterious linear optical losses over the HOM dip, becoming a good option for studying the elusive ETPA process. 

\section{\label{sec:Experimental-section}Experiment}
\begin{figure*}[h]
	\centering
	\includegraphics[width=\textwidth]{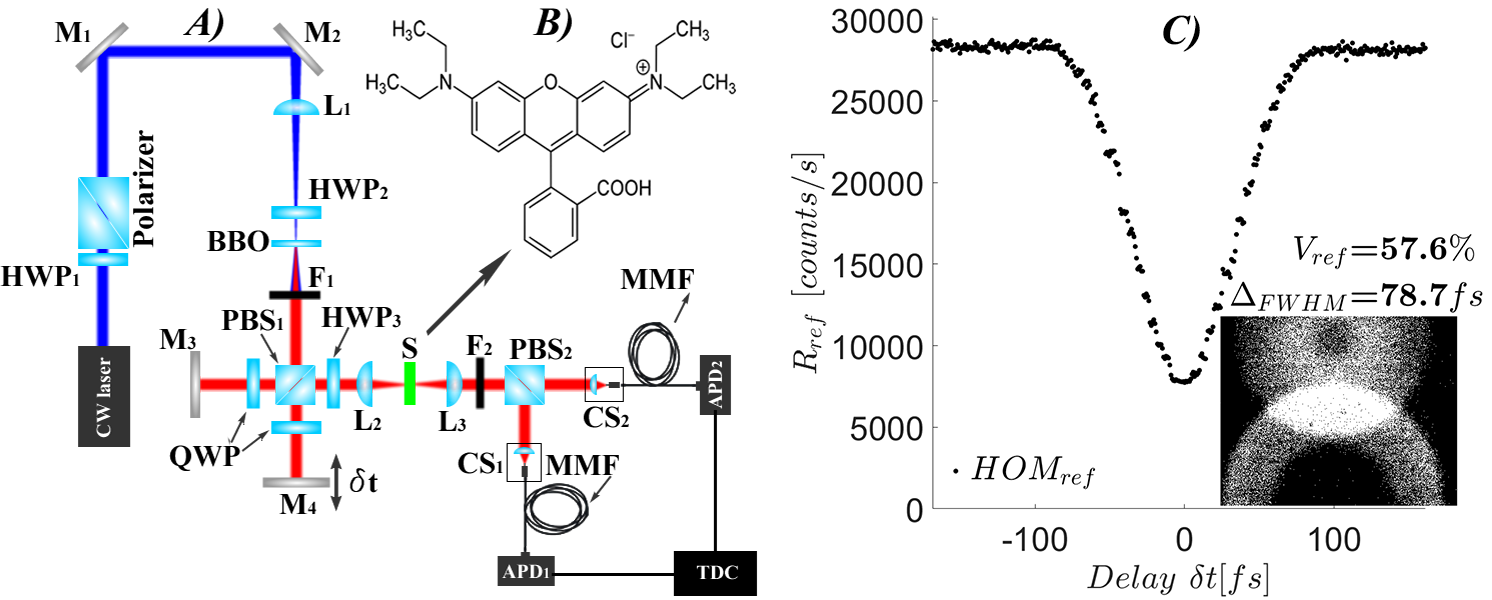}
	\caption{$A$) Experimental configuration for the transmission ETPA experiments using the HOM interferometer as a sensing device,  $B$) RhB  molecule, and $C$) calibration HOM dip ($HOM_{ref}$).} 
	\label{setup2}
\end{figure*}

We aim to perform ETPA transmission vs pump power\cite{VillabonaMonsalve2017} and transmission vs two-photon temporal delay experiments using a HOM interferometer in the CW excitation regime, obtaining information about the effects over the quantum state of type-II SPDC photon pairs, after interacting with a sample. The experimental setup is shown in Fig. \ref{setup2}$A$, where a CW laser (Crystalaser DL-405-100), with a center wavelength at $403nm$ and a $\sim 1nm$ bandwidth (FWHM), is focused by lens $L_1$ (focal length $f_1=500mm$) into a BBO crystal to generate collinear, cross-polarized and frequency non-degenerate Type-II SPDC photons pairs around $806nm$. To control the flux of photon pairs generated in the BBO crystal we use a half-wave plate ($HWP_{1}$) and a Glan-Thompson polarizer ($Polarizer$) after the CW laser, regulating the pumping power of the CW laser in a range from $0.25$mW to $43.9$mW. Then, the photon pairs are guided to a Michelson interferometer where a controllable temporal delay ($\delta t$) between them is introduced by a motorized stage. The delayed photon pairs are focused on the sample ($S$)  through lens $L_2$ (focal length $f_1=50mm$) and collected by $L_3$ with the same focal length. Additionally, to control the frequency and temporal indistinguishability of the photon pairs, an extra requirement to obtain an optimal HOM interference pattern at $PBS_{2}$ is to erase the polarization distinguishability, proper of a Type-II SPDC process. To do so, a half-waveplate ($HWP_{3}$) is introduced before $L_{2}$ in order to rotate $45^{\circ}$ the horizontal and vertical axis of polarization of the down-converted photons. An spectral filtering element, composed of a longpass filter (Thorlabs FELH0500) and a bandpass filter (Thorlbas FBH800-40), is used as $F_{1}$ after the BBO crystal to remove the residual $403$nm pump and control the bandwidth of the down-converted photons, and then $F_{2}$ (with the same configuration as $F_{1}$) is placed after the lens $L_{3}$ to eliminate a possible fluorescence residual signal from the sample. After the interaction of the photons with the sample, they arrive at the beam splitter ($PBS_{2}$) where the HOM interference occurs and then, the photons are collected with coupling systems ($CS_{1,2}$) into multi-mode fibers ($MMF$) to be detected by a pair of avalanche photodiodes (Excelitas SPCM-AQRH) ($APD_{1,2}$). Finally, the individual and coincidence counts ($CC$) between $APD_{1}$ and $APD_{2}$ are recorder by a time-to-digital converter module (ID Quantique id800) ($TDC$).

A key characteristic of the HOM dip is that it is defined by the quantum state of the photon pairs that interfere at the beam splitter\cite{Grice1997}, in the sense that any alteration in the state, due to their interaction with a sample, should be manifested as a change in the visibility ($V$) or the width of the dip ($\Delta t$)\cite{Freiman}. In this work, we focus on the changes introduced in the visibility since, as will be shown below, linear losses do not affect $V$ of the HOM dip, while it is expected that the nonlinear interaction is able to induce $V$ changes\cite{Freiman}. The HOM dip visibility is defined by
\begin{equation}
    V=\frac{R_{max}-R_{min}}{R_{max}+R_{min}},
    \label{VISIBILITY}
\end{equation}
where $R_{max}$, and $R_{min}$ are the maximum and minimum $CC$ rate, registered in our HOM apparatus at a temporal delay far from the center ($\delta t=167fs$) and at the center of the dip ($\delta t =0fs$), respectively. 

The experimental setup is calibrated by measuring the HOM two-photon interference pattern ($HOM_{ref}$) shown in Fig. \ref{setup2}$C$, obtained in free space propagation (no sample present). The resulting HOM dip has a temporal width of $\Delta t\sim 79fs$ (FWHM ), corresponding to a  bandwidth of the SPDC spectrum of $\Delta\lambda\sim24nm$ at $\lambda=806nm$. A good value of $V\sim 58\%$ was obtained, assuring the optimal alignment of the optical system \cite{Grice1997}. The spatial mode of Type II SPDC photons is shown in the inset of Fig. \ref{setup2}$C$. The region where the rings overlap comprises the entangled photons used to excite the sample.

\subsection{\label{sec:Results}Sample and transmission measurement procedure}
In this study we used as sample ($S$ as shown in Fig. \ref{setup2}$A$) an organic solution of the laser dye RhB ($\geq95\%$ purity, Sigma-Aldrich) dissolved in methanol at ten different concentrations: $0.01\mu$M, $0.1\mu$M, $1\mu$M, $0.01$mM, $0.1$mM, $1$mM, $4.5$mM, $10$mM, $58$mM, $100$mM. Some of these concentrations were chosen because they were employed in previous ETPA reports.  \cite{VillabonaMonsalve2017,VillabonaMonsalve2020,LandesTiemo2021,CoronaAquinoSamuel2022}. The solutions of RhB were contained in a 1 cm-thick quartz-cuvette.  

As a first step in the experimental procedure, the HOM dip of Fig. \ref{setup2}$C$ ($HOM_{ref}$) is used to set two temporal delay values of the photon pair: 1) the center of the HOM dip for zero delayed photons ($\delta t=0fs$), where a photon-pair flux density ($\phi$), calculated from the reference coincidence count rate ($R_{ref}$) and the focusing area ($A$) (beam waist $w_{0}=58\mu m$) of the down-converted photons, is on the order of $\phi=R_{ref}/A=7\times10^{7}~[photons~cm^{-2}~s^{-1}]$, and 2) far from the center of the dip, where a delay value of $\delta t=167fs$ is picked and where $\phi=R_{ref}/A=3\times10^{8}~[photons~cm^{-2}~s^{-1}]$. 

Then, we performed two different transmission experiments. In the first experiment, the transmission of the down-converted photons after interacting with the sample is registered as a function of the laser pump power. These measurements were performed by monitoring the $CC$ at $\delta t=0fs$ and $\delta t=167fs$ and with two conditions in which the cuvette was containing either the solvent alone  ($HOM_{sol}$) or the sample solution at different concentrations ($HOM_{sam}$). In the second experiment the $CC$ of the down-converted photons after interacting with the sample are monitored by continuously varying the temporal delay. In this case, the pump power was fixed to the maximum value used in the first experiment. We made this experiment for the cuvette containing only the solvent or the sample solution at different concentrations as well.

\section{\label{sec:Results and Discussion}Results and Discussion}
\subsection*{\label{sec:Transmission vs power and two-photon temporal delay}Transmission vs power and vs two-photon temporal delay}
\begin{figure*}[h]
	\centering
	\includegraphics[width=\textwidth]{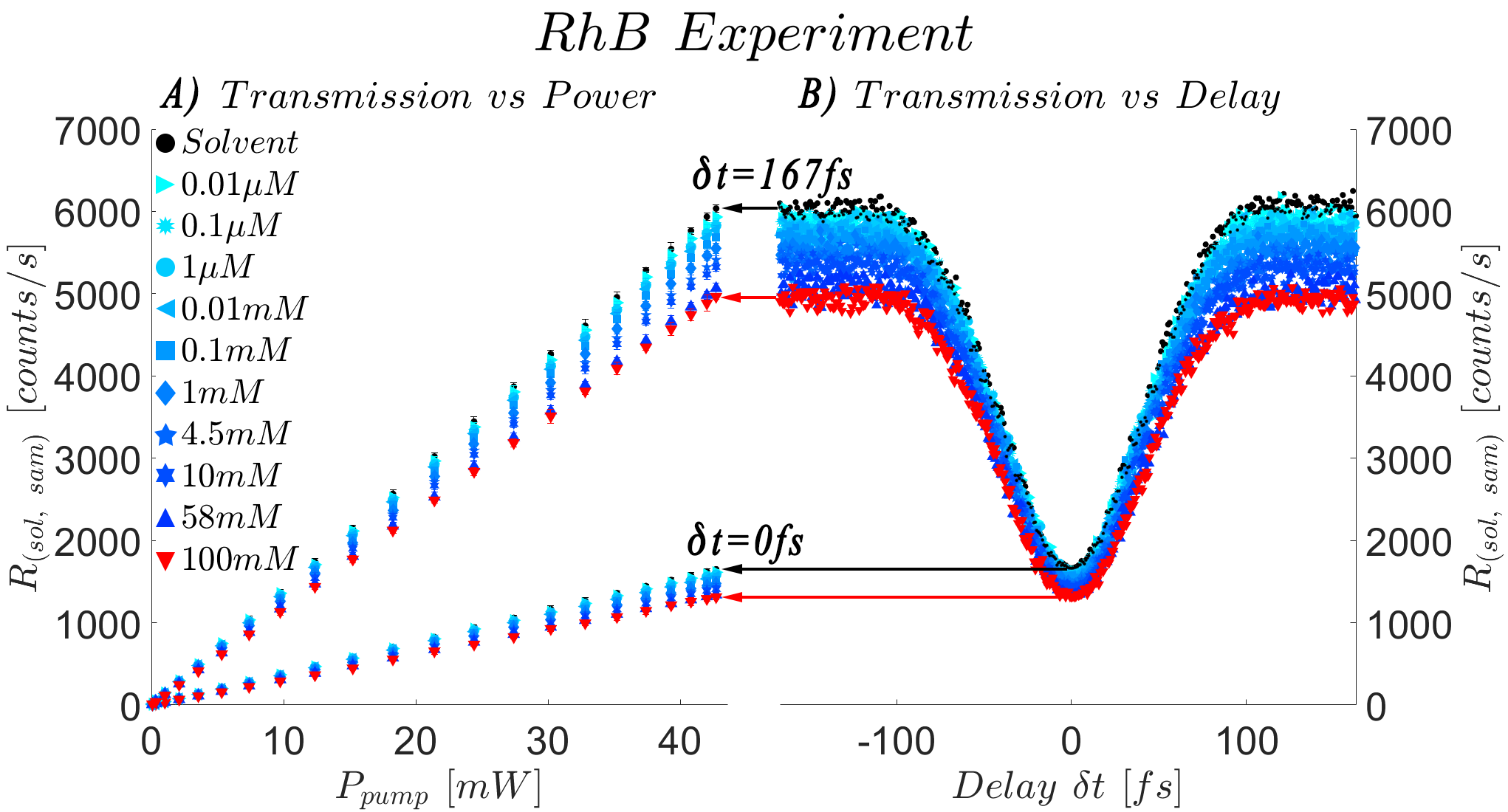}
	\caption{Transmission of solutions of the molecule RhB as a function of $A$) $P_{pump}$ and $B$) $\delta t$. In the case of transmission vs $\delta t$ the data was acquired by setting the pump power at $P_{pump}=43.9mW$ } 
	\label{RhB_experiment}
\end{figure*}

The measurements of transmission as a function of the pump power ($P_{pump}$) and transmission as a function of the two-photon temporal delay ($\delta t$) are presented in Fig. \ref{RhB_experiment}, where the rate of transmitted down-converted photons detected as $CC$ for the solvent and the samples are named $R_{sol}$ and $R_{sam}$, respectively, to be matched with tags used in previous works\cite{VillabonaMonsalve2017,CoronaAquinoSamuel2022}. The panel $A)$ of the figure shows $R_{(sol,~sam)}$ vs $P_{pump}$ for two specific delay configurations namely $\delta t=0fs$ and $\delta t = 167 fs$, while panel $B)$ shows $R_{(sol,~sam)}$ vs $\delta t$ for a fixed pump power of $P_{pump}=43.9mW$. The legend-color confections are selected to be: black-dots for solvent, red-dots for the most concentrated sample solution ($100$mM), and in a blue-cyan scale color the other sample concentrations in the range $0.01\mu$M - $58$mM.

In Fig. \ref{RhB_experiment} we clearly show the correspondence between the two transmission schemes when realizing $R_{(sol,~sam)}$ measurements, namely the transmission vs $P_{pump}$ in the standard configuration, or the transmission vs $\delta t$ of the HOM dip, as proposed in this study. Consequently, the difference in transmission between the solvent and sample ($R_{sol}-R_{sam}$) can be indistinctly quantified in the power measurements or in the temporal delay measurements. 

\begin{figure*}[h]
	\centering
	\includegraphics[width=\textwidth]{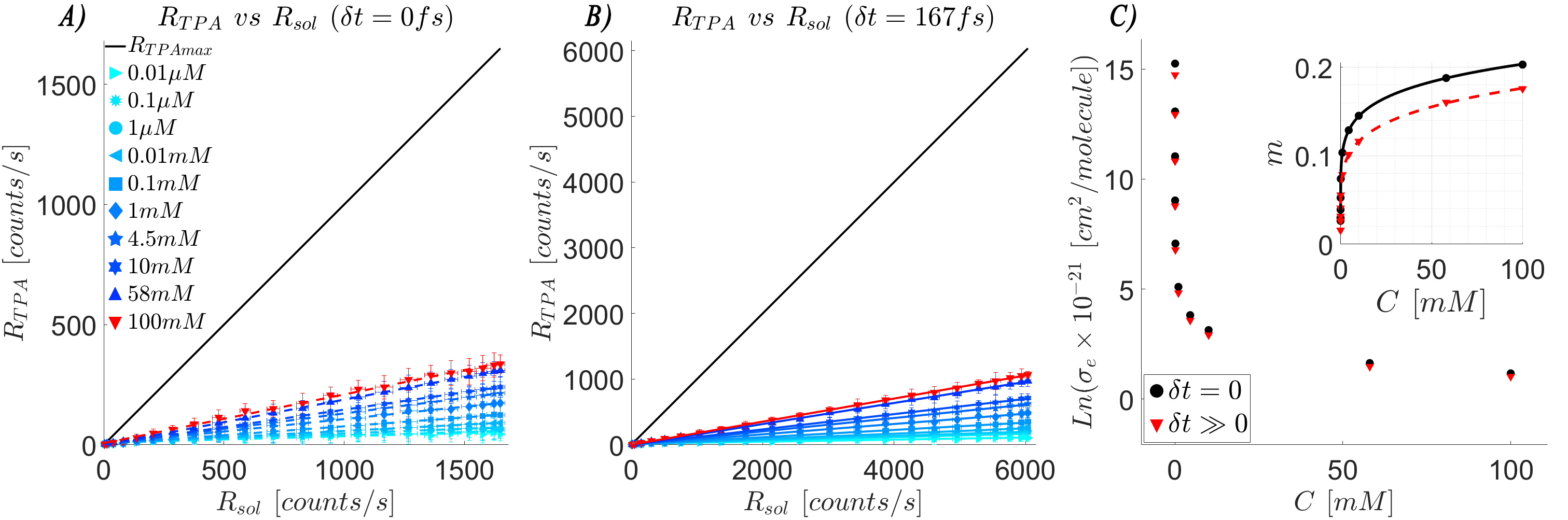}
	\caption{ETPA absorption rates of RhB for $A$) $\delta t= 0 fs$ and $B$) $\delta t =167 fs$. $C$)ETPA cross-section $\sigma_{e}$ (logarithmic scale) and slope $m$ (inset) as a function of concentration.} 
	\label{RTPA_experiment_sigma}
\end{figure*}

The difference $R_{sol}-R_{sam}$ is typically defined as $R_{TPA}$ \cite{VillabonaMonsalve2017}, but when this definition is employed special care must be taken to avoid misleading in the calculation of $\sigma_{e}$. For instance, when $R_{TPA}$  is computed directly from the data presented in Fig. \ref{RhB_experiment}, the largest values are obtained when the photons arrive at the sample with a temporal delay of $\delta t = 167 fs$, which is evidently an incorrect result since the simultaneous absorption of two photons must be maximum when the temporal delay between them is zero\cite{Parzuchowski2021,LandesTiemo2021,LandesTiemo2021_2}. Here it should be observed that the maximum delay of 167 $fs$ employed in these experiments is larger than the photon pair coherence time of $\sim$ 79 $fs$, as it can be seen in Fig. \ref{setup2}$C$. In the transmission vs $P_{pump}$ experiments, the quotient of $R_{TPA}$ with $R_{sol}$ must be computed for the whole pump-power interval, leading to  straight lines whose slopes are\cite{VillabonaMonsalve2017,CoronaAquinoSamuel2022}:
\begin{equation}
    m=\frac{R_{sol}-R_{sam}}{R_{sol}}=\frac{R_{TPA}}{R_{sol}},
    \label{SlopeValue}
\end{equation}
The quotient that defines $m$ considers the interferometric effect introduced by the HOM since it weighs the changes in the value of $R_{TPA}$ with respect to changes in the signal generated by the solvent alone, for different delay values. 
The lines obtained from the experimental data are shown in Fig. \ref{RTPA_experiment_sigma}$A$ for the case of $\delta t=0 fs$ and Fig. \ref{RTPA_experiment_sigma}$B$ for $\delta t=167 fs$. The black solid lines in these figures correspond to the hypothetical case of $m=1$, so they represent the maximum achievable level of ETPA, which occurs when the transmission of the sample is null ($R_{sam}=0$). The slope of the lines obtained experimentally are related to the $\sigma_{e}$ value by means of\cite{VillabonaMonsalve2017,CoronaAquinoSamuel2022}
\begin{equation}
    \sigma_{e}=\frac{m}{C}\frac{A}{V_{o}N_{A}},
    \label{SigmaValue}
\end{equation}
where $C$ is the concentration of the sample, $V_{o}$ and $A$ are the interaction volume and area, respectively, and $N_{A}$ is the Avogadro's number.  

The resulting values of $\sigma_{e}$ are presented in Fig. \ref{RTPA_experiment_sigma}$C$ (in logarithmic scale for clear visualization of data) as a function of sample concentration, while the inset of the figure shows the behavior of $m$. For illustration and discussion purposes, the results are presented for both delays, $\delta t =0fs$ and $\delta t =167fs$. Table \ref{TableRhB} summarizes these results and includes the corresponding uncertainties extracted directly from the fitting of experimental data in Fig.  \ref{RTPA_experiment_sigma}$A$ and Fig.  \ref{RTPA_experiment_sigma}$B$.

\begin{table}[]
\scalebox{0.9}{\begin{tabular}{|c|cc|cc|}
\hline
\multirow{2}{*}{$C[M/L]$}     & \multicolumn{2}{c|}{$\delta t=0fs$}                       & \multicolumn{2}{c|}{$\delta t=167fs$} \\ \cline{2-5} 
                       & \multicolumn{1}{c|}{$m$} & $\sigma_{e}$                & \multicolumn{1}{c|}{$m$}    & $\sigma_{e}$                  \\ \hline
$1\times10^{-8}$                & \multicolumn{1}{c|}{$0.026\pm0.002$}  & $(4.2\pm0.3)\times10^{-15}$               & \multicolumn{1}{c|}{$0.0158\pm0.0007$}    & $(2.5\pm0.1)\times10^{-15}$                 \\
$1\times10^{-7}$                & \multicolumn{1}{c|}{$0.030\pm0.002$}  & $(4.7\pm0.3)\times10^{-16}$               & \multicolumn{1}{c|}{$0.0263\pm0.0006$}    & $(4.2\pm0.1)\times10^{-16}$                 \\
$1\times10^{-6}$                & \multicolumn{1}{c|}{$0.039\pm0.002$}  & $(6.2\pm0.3)\times10^{-17}$               & \multicolumn{1}{c|}{$0.0315\pm0.0007$}    & $(5.0\pm0.1)\times10^{-17}$                 \\
$1\times10^{-5}$                & \multicolumn{1}{c|}{$0.0525\pm0.0007$} & $(4.2\pm0.1)\times10^{-18}$               & \multicolumn{1}{c|}{$0.0408\pm0.0008$}    & $(6.5\pm0.1)\times10^{-18}$                 \\
$1\times10^{-4}$                & \multicolumn{1}{c|}{$0.0740\pm0.0006$} & $(1.173\pm0.009)\times10^{-18}$             & \multicolumn{1}{c|}{$0.0556\pm0.0007$}    & $(8.8\pm0.1)\times10^{-19}$                 \\
$1\times10^{-3}$               & \multicolumn{1}{c|}{$0.103\pm0.001$}  & $(1.64\pm0.02)\times10^{-19}$              & \multicolumn{1}{c|}{$0.0786\pm0.0009$}    & $(1.25\pm0.01)\times10^{-19}$               \\
$4.5\times10^{-3}$               & \multicolumn{1}{c|}{$0.1289\pm0.0003$} & $(4.54\pm0.02)\times10^{-20}$              & \multicolumn{1}{c|}{$0.1015\pm0.0007$}    & $(3.58\pm0.03)\times10^{-20}$                \\
$1\times10^{-2}$                & \multicolumn{1}{c|}{$0.1452\pm0.0003$} & $(2.303\pm0.005)\times10^{-20}$            & \multicolumn{1}{c|}{$0.1163\pm0.0007$}    & $(1.84\pm0.01)\times10^{-20}$                \\
$5.8\times10^{-2}$                & \multicolumn{1}{c|}{$0.1879\pm0.0005$} & $(5.14\pm0.01)\times10^{-21}$              & \multicolumn{1}{c|}{$0.1605\pm0.0009$}    & $(4.39\pm0.03)\times10^{-21}$                \\
$1\times10^{-1}$                 & \multicolumn{1}{c|}{$0.203\pm0.003$}  & $(3.22\pm0.04)\times10^{-21}$              & \multicolumn{1}{c|}{$0.1758\pm0.0006$}    & $(2.79\pm0.01)\times10^{-21}$                \\ \hline
\end{tabular}}
\caption{Slope ($m$) values and ETPA cross-sections ($\sigma_{e}$) obtained from RhB solutions of different concentrations at two-photon delay ($\delta t$) configurations. The units of $\sigma_{e}$ are given in $[cm^{2}/molecule]$.}
\label{TableRhB}
\end{table}

Table \ref{TableRhB} shows that the difference between $\sigma_{e}$ measured with simultaneous photons ($\delta t =0fs$) and delayed photons ($\delta t =167fs$) is rather small. However, it must be noticed that in the case of $\delta t =167fs$ the two-photon absorption must be null since such temporal delay is longer than the two-photon coherence time. This fact clearly demonstrates that the calculated value of $\sigma_{e}$ at long delays can not be attributed to ETPA but must be the result from an artifact. Therefore, the measurements at $\delta t =0fs$ might also suffer from the same artifact, i.e., an optical loss emulating the ETPA process. In spite the transmission vs $P_{pump}$ experiments tend to produce untrue $\sigma_{e}$ values, it results illustrative to look at possible tendencies in the differences of $\sigma_{e}$ measured at $\delta t =0fs$ and $\delta t =167fs$. We perform an ANOVA test to evaluate if there is a significant statistical difference in the values of  $\sigma_{e}$ calculated at the two delay configurations across different concentrations. According to this test, the difference is statistically significant since $p\leq0.05$ (for data in Table \ref{TableRhB} $p = 0.05$ ). We believe that this apparent difference was biased by an artifact and not from ETPA, as discussed below considering the inconsistency of big variations of $\sigma_{e}$ across concentrations.  

The values of $\sigma_{e}$ presented in Table \ref{TableRhB} varied 6 orders of magnitude from the lowest to the highest concentration. This might represent an inconsistency since  $\sigma_{e}$ is an intensive (molecular) parameter that should not depend on concentration, as we can see by  differentiating Eq. (\ref{SigmaValue}) and noting that changes in concentration ($\delta C$) must be compensated by changes in the slope ($\delta m$), namely
\begin{equation}
\delta\sigma_{e}=\frac{A}{CVN_{A}}\left( \delta m-\frac{m}{C}\delta C \right)=0,
\label{sigmaVariation}
\end{equation}
where, 
\begin{equation}
    \delta m=\frac{m}{C}\delta C,
    \label{mCVariation}
\end{equation}
such that the quotient $\frac{m}{C}$ must be a constant. The inset of Fig. \ref{RTPA_experiment_sigma}$C$ presents the relation between $m$ and $C$, showing a clear non-constant relationship in our transmission vs $P_{pump}$ experiments. Further, it should be noticed that in Table \ref{TableRhB} there is a difference of 7 orders of magnitude in the concentrations while there is a difference of only 1 order of magnitude in the slopes, in contradiction with the expected behavior for ETPA where the slope should grow more rapidly in order to keep the ratio between $m$ and $C$ as a constant. This implies that the increasing difference between $R_{sam}$ and $R_{sol}$ as a function of concentration should be due to optical losses non-related to ETPA, for instance, it could be the effect of scattering, agglomeration and an increasing of the residual one-photon absorption occurring at high molecular concentration, as it has been suggested previously for the case of classical non-linear absorption\cite{AJAMI2015524}. 
 
We compare some of the $\sigma_{e}$ values previously reported in RhB with the values here obtained. For example, Villabona-Monsalve et al.\cite{VillabonaMonsalve2017} performed measurements for concentrations in the range 0.038 to 110 mM, obtaining $\sigma_{e}$ values from $4.20\times10^{-18}$ to $1.7\times10^{-20}$ $[cm^{2}/molecule]$; in particular, for the concentration of 4.5mM they reported $6.3\times10^{-20}$ $[cm^{2}/molecule]$, which is similar to the value of $4.54\times10^{-20}$ $[cm^{2}/molecule]$ we obtained at the same concentration when $\delta t=0fs$. In another interesting work\cite{CoronaAquinoSamuel2022}, Corona-Aquino et al., measured $\sigma_{e}$ values for various concentrations, and similarly to the present work, they also reported $\sigma_{e}$ at zero and temporal delays longer than the coherence time; for instance, for a sample of 4.5mM of concentration, they obtained $1.039\times10^{-20}$ and $1.041\times10^{-20}$ $[cm^{2}/molecule]$ for $\delta t=0fs$ and $\delta t=667fs$, respectively. They did not find any trend in the difference of $\sigma_{e}$ in both delay configurations. In similarity with our work, they also concluded that the no suppression of the apparent nonlinear absorption at long delays is due to artifacts. 

Lastly,  D. Tabakaev et al.\cite{DTabakaev2021} also presented estimations of $\sigma_{e}$ but for the case of Rhodamine 6G (a molecule of the Rhodamine family) using an experimental scheme based on transmission and fluorescence. At the 4.5mM concentration, the obtained value was $9.9\times10^{-22}$ $[cm^{2}/molecule]$, which is two orders of magnitude smaller than our result in RhB. In this cited work, and also in a work by T. Landes et al.\cite{LandesTiemo2021}, agglomeration effects were demonstrated in the fluorescence signal, which is an indication of optical losses not associated with ETPA that lead to overestimating the $\sigma_{e}$ values; consequently, authors proposed corrections resulting in significant smaller values of $\sigma_{e}$. 

Overall, all the cited works suggest that ETPA experiments based on transmission or fluorescence vs power schemes could have reported overstimulated values of $\sigma_{e}$ or effects unrelated to nonlinear absorption, due to unaccounted linear optical losses. 

To further investigate how the artifacts introduced by lineal optical losses in transmission experiments can be discriminated, we implemented another set of experiments in the scheme of transmission vs temporal delay, where a controllable source of linear optical losses is used, as explained below.

\subsection{\label{sec:Controllable optical losses introduced in ETPA transmission experiments}Controllable optical losses introduced in ETPA transmission experiments}    
To have a better understanding of how optical losses affect the ETPA transmission measurements, we designed two complementary experiments. The first experiment is the characterization of a two-photon interferogram comprising linear optical losses induced by different solvents. The second experiment allows us to introduce in a controllable way linear losses by adding nanospheres of silicon dioxide ($SiO_{2}$) in the solvent. In these tests, we performed transmission vs $\delta t$ measurements, which as we already showed, are fully equivalent to the transmission vs $P_{pump}$ measurements. In the first mentioned experiment, HOM dips are obtained for different solvents (methanol, ethanol, chloroform, ethylene glycol), and the corresponding linear losses are analyzed; in the second experiment, methanol is used as a solvent and the linear losses are studied as a function of the increasing concentration of dispersed nanoparticles. Since the nanoparticles do not have an absorption band in the wavelength range used in our tests\cite{Plautz2017,KousikDutta2006}, they effectively act as a light scattering medium producing controllable linear losses depending on the nanoparticle's concentration. 

\begin{figure*}[h]
	\centering
	\includegraphics[width=\textwidth]{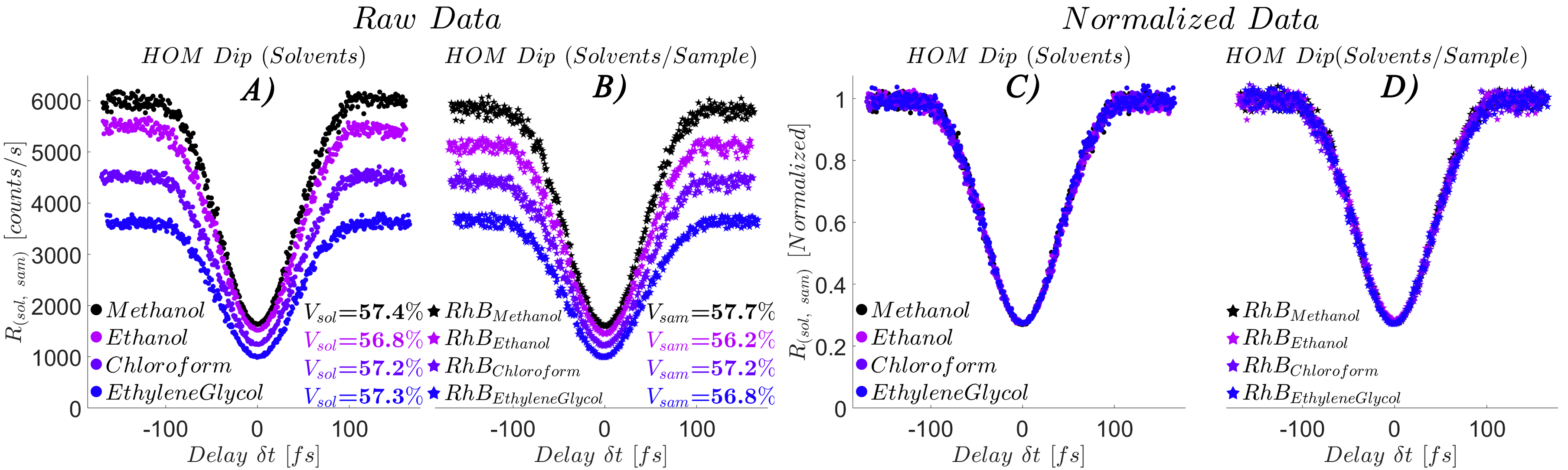}
	\caption{HOM interference experiment for solutions of RhB in different solvents. Panels A) and B) show raw data and panels C) and D) the corresponding normalization of panels A) and B. The HOM dips for solvents alone are shown in panels A) and C), while in panels B) and D) the HOM dips for solutions of RhB in four different solvents.} 
	\label{HOM_solvents_sample}
\end{figure*}

In Fig. \ref{HOM_solvents_sample} we present the obtained HOM dips for the four solvents. As a reference, the figure also includes the HOM dips from RhB dissolved at the concentration of 10mM in each of these solvents (solvents data are represented by dots while $RhB$ samples are represented by starts). This particular concentration was selected because is similar to those used by other authors to calculate $\sigma_{e}$ as it was discussed above.  
Various aspects can be extracted from these results; first, for each solvent different curves of $R_{sol}$ are registered. This is because each solvent produces a differentiated level of linear losses associated with Fresnel reflections, light scattering, etc. According to Fig. \ref{HOM_solvents_sample}$A$, the optical losses produce offsets between the HOM interferograms, where the smallest losses correspond to methanol, followed by ethanol, then chloroform, and finally ethylene glycol. Second, Fig. \ref{HOM_solvents_sample}$B$ shows the HOM dip of RhB in each solvent, where it can be noted that there is a small difference between the maximum value of $R_{sam}$ and its corresponding $R_{sol}$. Such differences are expected since RhB dissolved in each solvent leads to different levels of losses. Third, the visibility of the HOM dip for a RhB solution does not show a significant change compared with the  HOM dip of its corresponding solvent. In fact, and spite the different linear losses involved in each case, the temporal width ($\Delta t$) and depth of the dips remain unchanged as seen clearly by plotting normalized interferograms (see Fig. \ref{HOM_solvents_sample}$C$ and Fig. \ref{HOM_solvents_sample}$D$). Again, we perform an ANOVA test for the changes in visibility, obtaining a value of $p=0.56$, therefore the small differences between the visibility of $HOM_{sam}$ and $HOM_{sol}$ are not statistically significant, while they are rather associated to random experimental data fluctuations.  Table \ref{TableHOMsolventsSamples} summarizes these results. Here we recall our previous work on the interference of SPDC type-II photons\cite{Freiman}, where it was established that the ETPA activity introduces asymmetries in the joint spectral intensity (JSI) function, and in turn changes in the visibility of a HOM dip. Since the visibility of $HOM_{sam}$ shown in Fig. \ref{HOM_solvents_sample} is invariant with respect to $HOM_{sol}$, we conclude that there is no detection of ETPA signal. This outcome contrasts with that obtained in the standard transmission vs $P_{pump}$ experiment in which a value of $2.30\times10^{-20}$ $[cm^{2}/molecule]$ (see Table \ref{TableRhB} ) was calculated for the sample under test (concentration of 10mM in methanol).
\begin{table}[]
\begin{tabular}{|c|c|c|c|}
\hline
$Solvents$             & $V_{sol}[\%]$ & $R_{sol(max)}[counts/s]$ & $R_{sol(min)}[counts/s]$ \\ \hline
$Methanol$             & $57.4\pm1.5$      & $6035\pm101$           & $1635\pm101$           \\
$Ethanol$              & $56.8\pm1.6$      & $5506\pm96$           & $1516\pm96$           \\
$Chloroform$           & $57.2\pm1.6$      & $4552\pm81$           & $1240\pm82$           \\
$EthyleneGlycol$       & $57.3\pm1.7$      & $3636\pm68$           & $987\pm68$            \\ \hline
$Samples$              & $V_{sam}[\%]$ & $R_{sol(max)}[counts/s]$ & $R_{sam(min)}[counts/s]$ \\ \hline
$RhB_{Methanol}$       & $57.7\pm1.5$      & $5863\pm95$           & $1573\pm96$           \\
$RhB_{Ethanol}$        & $56.2\pm1.7$      & $5152\pm98$           & $1445\pm99$           \\
$RhB_{Chloroform}$     & $57.2\pm1.8$      & $4459\pm89$           & $1215\pm89$           \\
$RhB_{EthyleneGlycol}$ & $56.8\pm1.7$      & $3671\pm68$           & $1011\pm68$           \\ \hline
\end{tabular}
\caption{HOM dip properties for solutions of RhB in different solvents at 10mM concentration}
\label{TableHOMsolventsSamples}
\end{table}

\begin{figure*}[h]
	\centering
	\includegraphics[width=\textwidth]{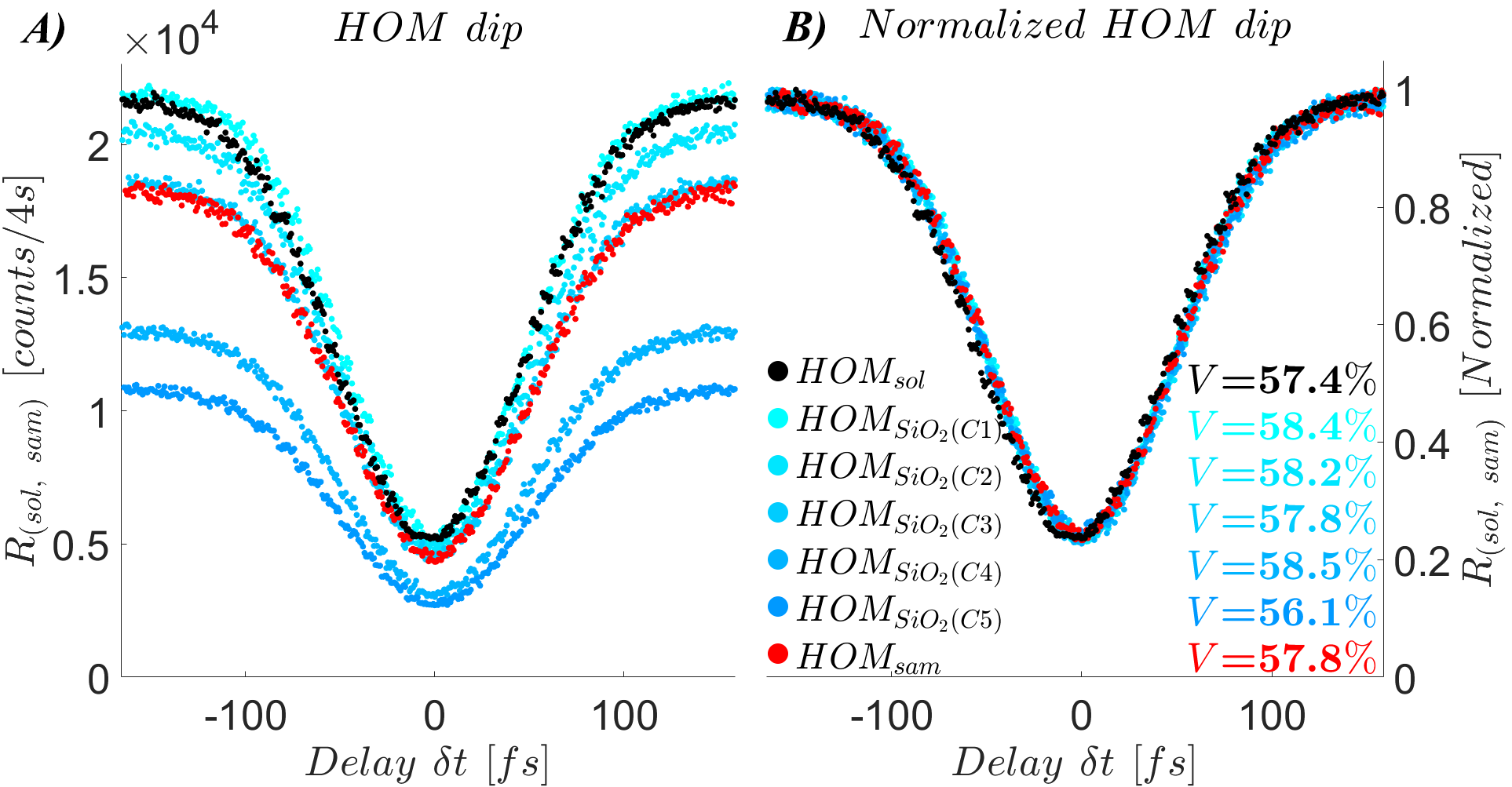}
	\caption{HOM interference experiment for controlled linear losses mechanism by adding $SiO2$ nanoparticles. $A)$ HOM dips for methanol solvent (black-dots), silicon nanoparticles suspended in methanol at different concentrations (cyan-dots), and RhB sample at the concentration of $100mM$(red-dots). $B)$ normalized HOM dips of the panel $A)$.} 
	\label{HOM_silicon}
\end{figure*}

In the second experiment, we implemented a method of controlled linear losses by adding silicon dioxide $(SiO_{2})$ nanoparticles (from Sigma Aldrich) of $10nm$ in diameter into the solvent (methanol) at 5 different concentrations: $C1=0.1$, $C2=5.2$, $C3=9.6$, $C4=15.3$, $C5=22$ $mg/10mL$. The nomenclature used for the HOM dips obtained with these samples is $HOM_{SiO_{2}(C)}$, where $C$ is the concentration of nanoparticles in the solvent. The HOM dips for nanoparticles solutions (cyan-dots), the HOM dip for solvent (black-dots) and the HOM dip for the most concentrated sample ($100mM$) of RhB ($HOM_{sam}$, red-dots) are shown in Fig. \ref{HOM_silicon}$A$ for data without any treatment and in Fig. \ref{HOM_silicon}$B$ for normalized data.

Figure \ref{HOM_silicon} displays the dips of $HOM_{SiO_{2}(C)}$, where higher concentration of nanoparticles implies stronger light scattering, and the subsequent reduction of $R_{sam}$. Notoriously, the strong transmission losses introduced by the scattering do not modify  the visibility of the interferograms, introducing only small random  fluctuations in acquired data. These experiments clearly show that linear losses in the transmission vs  $\delta t$ measurements can introduce noticeable offsets in the HOM dips without changing its inherent visibility, i.e., such losses do not alter the down-converted photon pair quantum state leaving invariant the dip visibility. 

Also, Fig. \ref{HOM_silicon} displays another interesting outcome: the plot of $HOM_{sam}$ at $100mM$ matches $HOM_{SiO_{2}(C3)}$. In line with our discussion, it follows that the effect of the light scattering introduced by nanoparticles of silica dissolved in methanol at the $C$3 concentration is equivalent to the transmission losses observed for RhB at $100mM$. The type of losses detected in RhB must be linear because any loss of photon pairs due to ETPA should be manifested as a change in the visibility of the normalized HOM dip. Indeed, the visibility is intrinsically related to the quantum state of the down-converted photons\cite{Grice1997,HaoLi2019}, while the absorption of entangled photons should modify the quantum state.\cite{Freiman} These observations, and the fact that in the transmission vs $P_{pump}$ experiments, there is no depletion of  $R_{TPA}$ signal even for temporal delay longer than the entanglement time, supports the conclusion that the calculated $\sigma_{e}$ values presented in Table \ref{TableRhB} resulted from artifacts produced by the sample which emulate the ETPA activity. 

\section{\label{sec:Conclusion}Conclusions}

The transmission of entangled photons from a nonlinear sample (RhB) was examined in two configurations, namely transmission vs $P_{pump}$ and transmission vs $\delta t$. In our experimental proposal, a HOM interferogram is registered after a pair of type II down-converted photons produced with a CW pump laser interact with the sample, with the premise that any effect that the sample produces over the two-photon quantum state should be observed as a change in the visibility of the HOM dip. 

Remarkably, the proposed scheme of transmission vs $\delta t$ based on the HOM interferogram results equivalent to the standard transmission vs $P_{pump}$ scheme and with this configuration we observed a persistence of $R_{TPA}$, even after delaying the photon pairs further than their coherence time, suggesting that the apparent ETPA signal detected in the transmission vs $P_{pump}$ experiments is generated by linear optical losses. Likewise, the variation of six orders of magnitude in the intensive parameter $\sigma_e$, for different RhB concentrations, also supports the conclusion that the detected $R_{TPA}$ signal can not be ascribed to an ETPA process.

Then, we performed a full analysis of the effects of controllable optical losses in the transmission vs $\delta t$, through the HOM interferogram. The obtained results show that linear losses from Fresnel reflections, scattering, aglomeration or residual one-photon absorption only reduce the counts in coincidence but without modifying the interferogram visibility. This is due to the fact that linear losses do not affect the quantum state of photon pairs. 

The invariant visibility of the HOM interferograms obtained for the RhB sample tested at different concentrations stongly suggests that ETPA activity was not detected in our study. Nevertheless, we show that the HOM dip is capable to account for the presence of linear losses affecting the ETPA signal, becoming a good strategy to study this elusive process.




\section{\label{sec:Associated Content}Associated Content}
\subsection{\label{sec:Data Availability Statement}Data Availability Statement}
        Data underlying the results presented in this paper are not publicly available at this time but may be obtained from the authors upon reasonable request.

\section{\label{sec:Author Information}Author Information}
    \subsection{\label{sec:Corresponding Authors}Corresponding Authors}
    \begin{itemize}
        \item \textbf{Freiman Triana-Arango} $-$ \textit{Centro de Investigaciones en Óptica A.C., A. P. 1-948, 37000 León, Gto, Mexico}; Email: freiman@cio.mx
        \item \textbf{Roberto Ramírez-Alarcón} $-$ \textit{Centro de Investigaciones en Óptica A.C., A. P. 1-948, 37000 León, Gto, Mexico}; Email: roberto.ramirez@cio.mx
        \item \textbf{Gabriel Ramos-Ortiz} $-$ \textit{Centro de Investigaciones en Óptica A.C., A. P. 1-948, 37000 León, Gto, Mexico}; Email: garamoso@cio.mx
    \end{itemize}

\section{\label{sec:Disclosures}Disclosures}
The authors declare no conflicts of interest.

\section{Acknowledgments}
We acknowledge support from CONACYT, Mexico. This work was supported by CONACYT, Mexico grant FORDECYT-PRONACES 217559.


\newpage

\section{\label{sec:TOC graphic}TOC graphic}

\begin{figure*}[h]
	\centering
	\includegraphics[width=\textwidth]{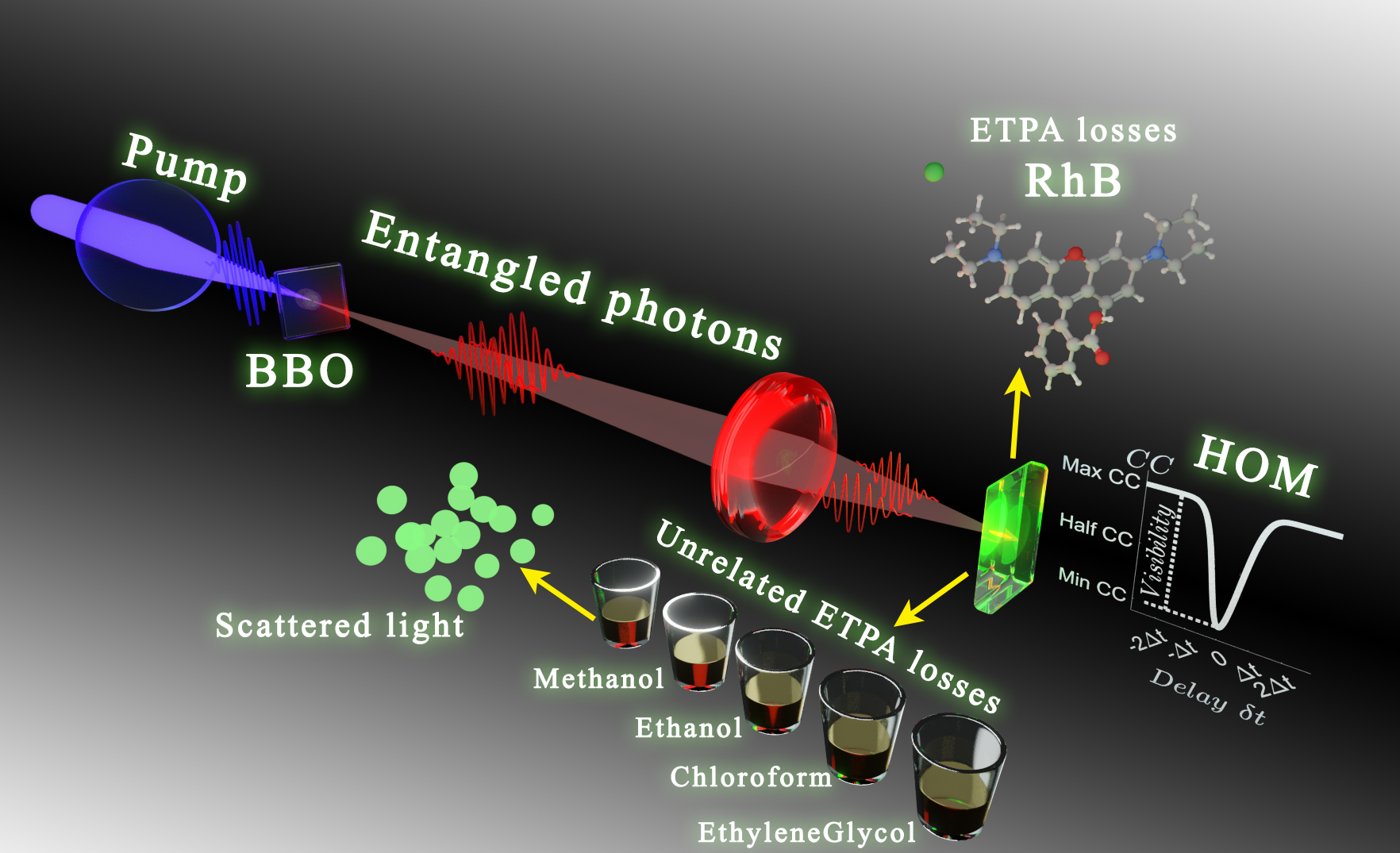}
	\label{GraphicAbstract }
\end{figure*}


\begin{thebibliography}{1}
	
\bibitem{MAIMAN1960} 
T.H. Maiman, Stimulated Optical Radiation in Ruby, Nature 187, 493-494 (1960).

\bibitem{Franken1961} 
P.A. Franken, A.E. Hill, C.W. Peters and G. Weinreich, Generation of Optical Harmonics, Phys. Rev. Lett. 7, 118-119 (1961).

\bibitem{Maker1962} 
P.D. Maker, R.W. Terhune, R.W. Nisenoff and C.M. Savage, Effects of Dispersion and Focusing on the Production of Optical Harmonics, Phys. Rev. Lett. 8, 21-22 (1962).

\bibitem{Armstrong1962} 
J.A. Armstrong, N. Bloembergen, J. Ducuing and P.S. Pershan, Interactions between Light Waves in a Nonlinear Dielectric, Phys. Rev. 127, 1918-1939 (1962).

\bibitem{Bloemberge1982} 
N. Bloembergen, Nonlinear optics and spectroscopy, Rev. Mod. Phys. 54, 685-695 (1982).

\bibitem{Shen1976} 
Y.R. Shen, Recent advances in nonlinear optics, Rev. Mod. Phys. 48, 1-32 (1976).

\bibitem{Chen1986} 
C. Chen and G. Liu, Recent Advances in Nonlinear Optical and Electro-Optical Materials, Annual Review of Materials Science 16, 203-243 (1986).

\bibitem{WDenk} 
W. Denk, J. H.Strickler and W. W.Webb, Two-photon laser scanning fluorescence microscopy, SCIENCE 248, 73 (1990).

\bibitem{Wei:07} 
WP. Wei, O. F. Tan, Y. Zhu and G. H. Duan, Axial superresolution of two-photon microfabrication, Appl. Opt. 46, 3694-3699 (2007).

\bibitem{Maker1965} 
P.D. Marker and R.W. Terhune, Study of Optical Effects Due to an Induced Polarization Third Order in the Electric Field Strength, Phys. Rev. 137, A801-A818 (1695).

\bibitem{MariacristinaRumi2010} 
M. Rumi and J.W. Perry, Two-photon absorption: an overview of measurements and principles, Adv. Opt. Photon. 4, 451-518 (2010).

\bibitem{GeaBanacloche1989} 
J. Gea-Banaloche, Two-photon absorption of nonclassical light, Phys. Rev. Lett. 62, 1603-1606 (1989).

\bibitem{Georgiades1995} 
N.Ph. Georgiades, E.S. Polzik, K. Edamatsu, H.J. Kimble and A.S. Parkins, Nonclassical Excitation for Atoms in a Squeezed Vacuum, Phys. Rev. Lett. 75, 3426-3429 (1995).

\bibitem{JanPerina1998} 
J. Perina, B.E.A. Saleh, and M.C. Teich, Multiphoton absorption cross section and virtual-state spectroscopy for the entangled n-photon state, Phys. Rev. A 57, 3972 (1998).

\bibitem{Fei1997} 
H.B. Fei, B.M. Jost, S. Popescu, B.E.A. Saleh, and
M.C. Teich, Entanglement-induced two-photon transparency, Phys. Rev. Lett. 78, 1679-1682 (1997).

\bibitem{Saleh1998} 
B.E.A. Saleh, B.M. Jost, H.B. Fei, M.C. Teich, Entangled-Photon Virtual-State Spectroscopy, Phys. Rev. Lett. 80, 3483-3486 (1998).

\bibitem{KOJIMA2004323} 
J. Kojima and Q. Nguyen, Entangled biphoton virtual-state spectroscopy of the $A^{2}\Sigma^{+}-X^{2}\Pi$ system of OH, Chem. Phys. Lett. 396, 323-328 (2004).

\bibitem{RobertoLeon2013} 
R.J. León-Montiel, J. Svozilík, L.J. Salazar-Serrano and J.P. Torres, Role of the spectral shape of quantum correlations in two-photon virtual-state spectroscopy, New J. Phys 15, 053023 (2013).

\bibitem{Muthukrishnan2004} 
A. Muthukrishnan, G.S. Agarwal, M.O. Scully, Inducing Disallowed Two-Atom Transitions with Temporally Entangled Photons, Phys. Rev. Lett. 93, 093002 (2004).

\bibitem{SVOZILIK201854} 
J. Svozilík, J. Peřina and R.J. León-Montiel, Two-photon absorption spectroscopy using intense phase-chirped entangled beams, Chemical Physics 510, 54-59 (2018).

\bibitem{Schlawin2018} 
F. Schlawin, K. Dorfman and M. Shaul, Entangled Two-Photon Absorption Spectroscopy, Accounts of Chemical Research 51, 2207-2214 (2018).

\bibitem{RobertoLeon2019} 
R.J. León-Montiel, J. Svozilík, J.P. Torres and A.B. U'Ren, Temperature-Controlled Entangled-Photon Absorption Spectroscopy, Phys. Rev. Lett. 123, 023601 (2019).

\bibitem{Mertenskotter:21} 
L. Mertenskötter, K. Busch, and R.J. León-Montiel, Entangled two-photon absorption spectroscopy with varying pump wavelengths, J. Opt. Soc. Am. B 38, c63-c68 (2021).

\bibitem{Javanainen1990} 
J. Javanainen, P.L. Gould, Linear intensity dependence of a two-photon transition rate, Phys. Rev. A 41, 5088-5091 (1990).

\bibitem{DongIkLee2007} 
D. Lee and T. Goodson III, Quantum spectroscopy of an organic material utilizing entangled and correlated photon pairs, Proc. SPIE 6653, Linear and Nonlinear Optics of Organic Materials VII, 66530V (2007).

\bibitem{Sperber1986} 
P. Sperber and A. Penzkofer, S0-Sn two-photon absorption dynamics of rhodamine dyes, Optical and Quantum Electronics 18, 381-401 (1986).

\bibitem{Xu:96} 
C. Xu and W.W. Webb, Measurement of two-photon excitation cross sections of molecular fluorophores with data from 690 to 1050 nm, J. Opt. Soc. Am. B 13, 481-491 (1996).

\bibitem{Makarov:08} 
N.S. Makarov, M. Drobizhev and A. Rebane, Two-photon absorption standards in the 550--1600 nm excitation wavelength range, Opt. Express 16, 4029-4047 (2008).

\bibitem{GhoshR1986} 
R. Ghosh, C.K. Hong, Z.Y. Ou and L. Mandel, Interference of two photons in parametric down conversion, Phys. Rev. A 34, 3962-3968 (1986).

\bibitem{KwiatPaulG1995} 
P.G. Kwiat, K. Mattle, H. Weinfurter, A. Zeilinger, A. V. Sergienko, and Y. Shih, New High-Intensity Source of Polarization-Entangled Photon Pairs, Phys. Rev. Lett. 75, 4337-4341 (1995).

\bibitem{Dayan2005} 
B. Dayan, A. Pe’er, A.A. Friesem, and Y. Silberberg, Nonlinear Interactions with an Ultrahigh Flux of Broadband Entangled Photons, Phys. Rev. Lett. 94, 043602 (2005).

\bibitem{LeeDongIk2006} 
D. Lee and T. Goodson, Entangled Photon Absorption in an Organic Porphyrin Dendrimer, The Journal of Physical Chemistry B 110, 25582-25585 (2006).

\bibitem{HarphamMichaelR2009} 
M.R. Harpham, Ö. Süzer, C. Ma, P. Bäuerle, and T. Goodson, III, Thiophene Dendrimers as Entangled Photon Sensor Materials, Journal of the American Chemical Society 131, 973-979 (2009).

\bibitem{Dayan2004} 
B. Dayan, A. Pe’er, A. A. Friesem, and Y. Silberberg, Two photon absorption and coherent control with broadband down-converted light, Phys. Rev. Lett. 93, 023005 (2004).

\bibitem{AviPeer2005} 
A. Pe’er, B. Dayan, A. A. Friesem, and Y. Silberberg, Temporal Shaping of Entangled Photons, Phys. Rev. Lett. 94, 073601 (2005).

\bibitem{Dayan2007} 
B. Dayan, Theory of two-photon interactions with broadband down-converted light and entangled photons, Phys. Rev. A 76, 043813 (2007).

\bibitem{YouHao2009} 
H. You, S. M. Hendrickson, and J. D. Franson, Enhanced two-photon absorption using entangled states and small mode volumes, Phys. Rev. A 80, 043823 (2009).

\bibitem{DanielKeefer2022} 
B. Gu, D. Keefer and S. Mukamel, Wave Packet Control and Simulation Protocol for Entangled Two-Photon Absorption of Molecules, Journal of Chemical Theory and Computation 18, 406-414 (2022).

\bibitem{VillabonaMonsalve2017} 
J. P. Villabona-Monsalve, O. Calder\'on-Losada, M. Nu\~{n}ez Portela, and A. Valencia, Entangled two photon absorption cross section on the 808 nm region for the common dyes zinc tetraphenylporphyrin and rhodamine b, The Journal of Physical Chemistry A 121, 7869 (2017).

\bibitem{VillabonaMonsalve2020} 
J.P. Villabona-Monsalve, R.K. Burdick, and T. Goodson III, Measurements of Entangled Two-Photon Absorption in Organic Molecules with CW-Pumped Type-I Spontaneous Parametric Down-Conversion, The Journal of Physical Chemistry C 124, 24526-24532 (2020).

\bibitem{UptonL2013} 
L. Upton, M. Harpham, O. Suzer, M. Richter, S. Mukamel, and T. Goodson III, Optically Excited Entangled States in Organic Molecules Illuminate the Dark, The Journal of Physical Chemistry Letters 4, 2046-2052 (2013).

\bibitem{VillabonaMonsalve2018} 
J.P. Villabona-Monsalve, O. Varnavski, B.A. Palfey, and T. Goodson III, Two-Photon Excitation of Flavins and Flavoproteins with Classical and Quantum Light, Journal of the American Chemical Society 140, 14562-14566 (2018).

\bibitem{VarnavskiOleg2020} 
O. Varnavski and T. Goodson III, Two-Photon Fluorescence Microscopy at Extremely Low Excitation Intensity: The Power of Quantum Correlations, Journal of the American Chemical Society 142, 12966-12975 (2020).

\bibitem{EshunAudrey2022} 
A. Eshun, O. Varnavski, J.P. Villabona-Monsalve, R.K. Burdick, and T. Goodson III, Entangled Photon Spectroscopy, Accounts of Chemical Research 55, 991-1003 (2022).

\bibitem{Mikhaylov2022} 
A. Mikhaylov, R.N. Wilson, K.M. Parzuchowski, M.D. Mazurek, C.H. Camp Jr., M.J. Stevens and R. Jimenez, Hot-Band Absorption Can Mimic Entangled Two-Photon Absorption, The Journal of Physical Chemistry Letters 13, 1489-1493 (2022).

\bibitem{BrycePHickam2022} 
B.P. Hickam, M.He, N.Harper, S.Szoke, and S.K. Cushing, Single-Photon Scattering Can Account for the Discrepancies among Entangled Two-Photon Measurement Techniques, The Journal of Physical Chemistry Letters 13, 4934-4940 (2022).

\bibitem{Parzuchowski2021} 
K.M. Parzuchowski, A.Mikhaylov, M.D. Mazurek, R.N. Wilson, D.J. Lum, T.Gerrits, C.H. Camp, Jr., M.J. Stevens and R. Jimenez, Setting Bounds on Entangled Two-Photon Absorption Cross Sections in Common Fluorophores, Phys. Rev. Applied 15, 044012 (2021).

\bibitem{LandesTiemo2021} 
T. Landes, M. Allgaier, S. Merkouche, B. J. Smith, A. H. Marcus, and M. G. Raymer, Experimental feasibility of molecular two-photon absorption with isolated time-frequency-entangled photon pairs, Phys. Rev. Research 3, 033154 (2021).

\bibitem{LandesTiemo2021_2} 
T. Landes, M.G. Raymer, M. Allgaier, S. Merkouche, B.J. Smith and A.H. Marcus, Quantifying the enhancement of two-photon absorption due to spectral-temporal entanglement, Opt. Express, 29, 20022-20033 (2021).

\bibitem{CoronaAquinoSamuel2022} 
S. Corona-Aquino, O. Calderón-Losada, M.Y. Li-Gómez,H. Cruz-Ramírez, V. Alvarez-Venicio, M. del Pilar Carreón-Castro, R.J. León-Montiel, and A.B. U’Ren, Experimental study on the eﬀects of photon-pair temporal correlations in entangled two-photon absorption, The Journal of Physical Chemistry A 126, 2185-2195 (2022).

\bibitem{Freiman} 
F. Triana-Arango, G. Ramos-Ortiz and R. Ramírez-Alarcón, Spectral Considerations of Entangled Two-Photon Absorption Effects in Hong–Ou–Mandel Interference Experiments, The Journal of Physical Chemistry A 127, 2608-2617 (2022).

\bibitem{Grice1997} 
W.P. Grice and I.A. Walmsley, Spectral information and distinguishability in type-II down-conversion with a broadband pump, Phys. Rev. A 56, 1627-1634 (1997).

\bibitem{AudreyEshun2021} 
A. Eshun, B. Gu, O. Varnavski, S. Asban, K. E. Dorf-
man, S. Mukamel, and T. Goodson III, Investigations of molecular optical properties using quantum light and hong–ou–mandel interferometer, The Journal of the American Chemical Society 143, 9070-9081 (2021).

\bibitem{KonstantinEDorfman2021} 
K.E. Dorfman, S. Asban, B. Gu and S. Mukamel, Hong-Ou-Mandel interferometry and spectroscopy using entangled photons, Communications Physics 4, 24526 (2021).

\bibitem{YuanyuanChen2022} 
Y.Chen, Q. Shen, S. Luo, L. Zhang, Z. Chen and L. Chen, Entanglement-Assisted Absorption Spectroscopy by Hong-Ou-Mandel Interference, Phys. Rev. Applied 17, 014010 (2022).

\bibitem{NicolasFabre2022} 
N. Fabre, Spectral single photons characterization using generalized Hong–Ou–Mandel interferometry, Journal of Modern Optics 69, 653-664 (2022).

\bibitem{AJAMI2015524} 
A. Ajami, P. Gruber, M. Tromayer, W. Husinsky, J. Stampfl, R. Liska, A. Ovsianikov, Evidence of concentration dependence of the two-photon absorption cross section: Determining the “true” cross section value, Optical Materials 47, 524-529 (2015).

\bibitem{HaoLi2019} 
H. Li, A. Piryatinski, A.R. Srimath-Kandada, C. Silva and E.R Bittner, Photon entanglement entropy as a probe of many-body correlations and fluctuations, The Journal of chemical physics 150, 184106 (2019).

\bibitem{Plautz2017} 
G.L. Plautz, I.L. Graff, W.H. Schreiner and A.G. Bezerra, Evolution of size distribution, optical properties, and structure of Si nanoparticles obtained by laser-assisted fragmentation, Applied Physics A 123, 359 (2017).

\bibitem{KousikDutta2006} 
K. Dutta and S.K. De, Transport and optical properties of SiO2–polypyrrole nanocomposites, Solid State Communications 140, 167-171 (2006).

\bibitem{DTabakaev2021} 
D. Tabakaev, M. Montagnese, G. Haack, L. Bonacina,
J.P. Wolf, H. Zbinden, and R. T. Thew, Energy-time entangled two-photon molecular absorption, Phys. Rev. A 103, 033701 (2021).


\end{thebibliography}
\end{document}